\documentclass[journal]{IEEEtran}

\usepackage{cite}
\usepackage{amssymb}
\usepackage[pdftex]{graphicx}
\usepackage[cmex10]{amsmath}
\usepackage{array}
\usepackage[caption=false,font=footnotesize]{subfig}
\usepackage{hyperref}
\usepackage{algorithm}
\usepackage{algpseudocode}

\begin{document}

\title{On the Unique Determination of Modal Multiconductor Transmission-Line Properties}

\author{Stuart~Barth,~\IEEEmembership{Graduate Student Member,~IEEE,}
        and~Ashwin~K.~Iyer,~\IEEEmembership{Senior Member,~IEEE}
\thanks{S. Barth and A. K. Iyer are with the Department
of Electrical and Computer Engineering, University of Alberta, Edmonton,
AB, Canada. e-mail: sbarth@ualberta.ca}
}

\markboth{Journal of \LaTeX\ Class Files,~Vol.~13, No.~9, September~2014}%
{Shell \MakeLowercase{\textit{et al.}}: Bare Demo of IEEEtran.cls for Journals}

\maketitle

\begin{abstract}
Some modal (or \emph{decoupled}) transmission-line properties such as per-unit-length impedance, admittance, or characteristic impedance have long been held to be, in general, non-unique. This ambiguity arises from the nature of the similarity transformations used to relate the terminal and modal domains, for which the voltage transformation matrix has been shown to be only loosely related to the corresponding current transformation matrix. Modern methods have attempted to relate the two, but these relations typically rely on arbitrary normalizations, leading to strictly incorrect and/or non-unique results. This work introduces relations between the two transformations, derived from the physical equivalence of total power and currents between the two domains, by which the transformation matrices can be unambiguously related to each other, and the modal properties uniquely solved. This technique allows for the correct extraction of the modal transmission-line properties for any arbitrary system of conductors. Multiple examples are studied to validate the proposed solution process.
\end{abstract}


\section{Introduction}

\IEEEPARstart{T}{ransmission}-line (TL) theory is a powerful concept which simplifies the fields of transverse electromagnetic (TEM) modes and allows them to be expressed as circuit quantities, making it a critical tool for high-accuracy circuit design. Basic TL systems contain only two conductors, whereas generalized \emph{multiconductor} TL (MTL) systems may contain three or more conductors. The two common representations (or ``domains") of MTLs are as follows:
\newline
\begin{enumerate}
\item The \emph{terminal} domain (also referred to as the \emph{natural} domain), in which the various TL parameters are defined between each conductor and a pre-selected reference conductor. These parameters are expressed as matrices which are (generally) fully populated, which implies that the terminals are coupled to each other.
\newline
\item The \emph{modal} domain (also referred to as the \emph{diagonalized} or \emph{decoupled} domain), in which the TL's properties are given in terms of the TL's characteristic modes and expressed as diagonal matrices. The diagonality of the matrices implies that the solutions are isolated from one another, and due to this fact, each mode can be expressed by a simple two-conductor TL model.
\end{enumerate}

A process exists by which the terminal-domain parameters may be transformed into the modal domain, and vice-versa \cite{ref:Paul_Book, ref:Faria_Book,ref:Marx_Paper1}. This process, known as diagonalization, results in the solution of two \emph{transformation} matrices, which are responsible for transforming the currents and voltages (and by extension, other parameters) between the two domains. These matrices are solved independently, which is the source of some ambiguity when they are used together. The specific process of, for example, the diagonalization of the propagation constant requires that only one of the transformation matrices be used and solved; however, the process for the diagonalization of some properties such as the characteristic impedance involves both the transformation matrices. This fact has historically resulted in the conclusion that some modal properties such as characteristic impedance are generally ambiguous when transformed from the terminal domain (although other methods may of course be used to obtain the correct values, such as derivation from known modal field quantities \cite{ref:Conventional_Zc_1,ref:Conventional_Zc_2}).

Over roughly the last two decades, solutions have been proposed in attempts to overcome this ambiguity \cite{ref:Paul_Paper1,ref:Faria_Paper1,ref:Normalized_Eigenvalues1, ref:Normalized_Eigenvalues2}, and present similar processes to what will be introduced in this work, but generally either utilize ambiguous normalizations that lack physical bases, or produce non-unique results.

This work demonstrates that any such ambiguities can be resolved by noting that physical quantities -- specifically, total power and current -- must be equivalent under both forms of representation. These physical facts are used together  to constrain the spectrum of possible mathematical solutions and produce unique modal results. While attempts have been made to equate total real power in both domains \cite{ref:Paul_Paper1}, it will be shown that the consideration of total power (i.e., including the reactive component) is required in order to produce a unique solution. Additionally, it can be demonstrated that given the equivalence of these properties in the two domains, the transformation matrices are required to be real.

The layout of this document is as follows: the analytical process of the terminal-to-modal transformation is detailed in Sec. \ref{sec:Theory}, along with the mathematical description of the ambiguity in the transforms' relations, a brief overview of previously proposed solutions, and its resolution. Sec. \ref{sec:Examples} details the data obtained for a number of TLs and demonstrates how the correct modal characteristic impedances may be directly computed using only the terminal domain per-unit-length inductance and capacitance matrices, even in the presence of extreme loss.

\section{Theory}
\label{sec:Theory}

\subsection{Diagonalization Procedure}
\label{sec:Decoupling_Procedure}
The analysis of TL systems typically begins with extracting their per-unit-length inductance and capacitance matrices ($\left[L\right]$ and $\left[C\right]$, respectively). These matrices detail the terminal-domain values, but for clarity in the following analysis, the subscripts $T$ and $M$ will be used to indicate which domain (terminal or modal) is being used in each matrix, e.g., $\left[L_T\right]$ or $\left[C_M\right]$.

The transformation between the two domains makes use of similarity transforms, specified in terms of currents and voltages, and which are expressed as (nonsingular) matrices \cite{ref:Paul_Book,ref:Faria_Book}. These transformations are defined as
\begin{subequations}
\begin{equation} \label{eq:Tv_Def}
\vec{V}_T = \left[T_V\right] \vec{V}_M
\end{equation}
\begin{equation} \label{eq:Ti_Def}
\vec{I}_T = \left[T_I\right] \vec{I}_M
\end{equation}
\end{subequations}
such that each column of $\left[T_I\right]$ and $\left[T_V\right]$ corresponds to a particular mode, and for which the entries describe the \emph{relative} (i.e., with respect to the modal quantities) currents and voltages on each conductor for that mode. The TL wave equations for propagation along $z$ in the terminal domain can be expressed as \cite{ref:Paul_Book}
\begin{equation} \label{eq:ZY_Terminal_Equations}
\setlength{\arraycolsep}{1pt}
\frac{\partial}{\partial z}
\left[ \begin{array}{cc}
\vec{V}_T \\
\vec{I}_T \\
\end{array} \right] = 
\left[ \begin{array}{cc}
\left[0\right] & -\left[Z_T\right]\\
-\left[Y_T\right] & \left[0\right]\\
\end{array} \right]
\left[ \begin{array}{cc}
\vec{V}_T \\
\vec{I}_T \\
\end{array} \right]
\end{equation}
where $\left[Z_T\right]$ and $\left[Y_T\right]$ are the per-unit-length impedance and admittance, respectively, in the terminal domain. The modal-domain equations have the same form; that is
\begin{equation} \label{eq:ZY_Modal_Equations}
\setlength{\arraycolsep}{1pt}
\frac{\partial}{\partial z}
\left[ \begin{array}{cc}
\vec{V}_M \\
\vec{I}_M \\
\end{array} \right] = 
\left[ \begin{array}{cc}
\left[0\right] & -\left[Z_M\right]\\
-\left[Y_M\right] & \left[0\right]\\
\end{array} \right]
\left[ \begin{array}{cc}
\vec{V}_M \\
\vec{I}_M \\
\end{array} \right]
\end{equation}
Inserting (\ref{eq:Tv_Def}) and (\ref{eq:Ti_Def}) into (\ref{eq:ZY_Terminal_Equations}) yields
\begin{equation} \label{eq:ZY_Modal_Equations_Solution1}
\setlength{\arraycolsep}{1pt}
\begin{gathered}
\frac{\partial}{\partial z}
\left[ \begin{array}{cc}
\left[T_V\right] & \left[0\right]\\
\left[0\right] & \left[T_I\right]\\
\end{array} \right]
\left[ \begin{array}{cc}
\vec{V}_M\\
\vec{I}_M\\
\end{array} \right] = \\
\left[ \begin{array}{cc}
\left[0\right] & -\left[Z_T\right]\\
-\left[Y_T\right] & \left[0\right]\\
\end{array} \right]
\left[ \begin{array}{cc}
\left[T_V\right] & \left[0\right]\\
\left[0\right] & \left[T_I\right]\\
\end{array} \right]
\left[ \begin{array}{cc}
\vec{V}_M \\
\vec{I}_M \\
\end{array} \right]
\end{gathered}
\end{equation}
which can be re-arranged and compared to (\ref{eq:ZY_Modal_Equations}) to give the relations
\begin{subequations}
\begin{equation} \label{eq:Z_Transform}
\left[Z_M\right] = \left[T_V\right]^{\text{-}1} \left[Z_T\right] \left[T_I\right]
\end{equation}
\begin{equation} \label{eq:Y_Transform}
\left[Y_M\right] = \left[T_I\right]^{\text{-}1} \left[Y_T\right] \left[T_V\right]
\end{equation}
\end{subequations}
Using these definitions, the propagation constants and characteristic impedances can be expressed as
\begin{subequations}
\begin{equation} \label{eq:G_Definition_1}
\left[\gamma_M\right]^2 = \left[Z_M\right]\left[Y_M\right] = \left[T_V\right]^{\text{-}1} \left[Z_T\right] \left[Y_T\right] \left[T_V\right]
\end{equation}
\begin{equation} \label{eq:G_Definition_2}
\left[\gamma_M\right]^2 = \left[Y_M\right]\left[Z_M\right] = \left[T_I\right]^{\text{-}1} \left[Y_T\right] \left[Z_T\right] \left[T_I\right]
\end{equation}
\end{subequations}
\begin{subequations}
\begin{equation} \label{eq:Zc_Definition_1}
\small
\left[{Z_c}_M\right] = \left(\left[Z_M\right]\left[Y_M\right]^{\text{-}1}\right)^{\frac{1}{2}} = \left(\left[T_V\right]^{\text{-}1} \left[Z_T\right] \left[T_I\right] \left[T_V\right]^{\text{-}1} \left[Y_T\right]^{\text{-}1} \left[T_I\right] \right)^{\frac{1}{2}}
\end{equation}
\begin{equation} \label{eq:Zc_Definition_2}
\small
\left[{Z_c}_M\right] = \left(\left[Y_M\right]^{\text{-}1}\left[Z_M\right]\right)^{\frac{1}{2}} = \left(\left[T_V\right]^{\text{-}1} \left[Y_T\right]^{\text{-}1} \left[T_I\right] \left[T_V\right]^{\text{-}1} \left[Z_T\right] \left[T_I\right] \right)^{\frac{1}{2}}
\end{equation}
\end{subequations}
where $\left[\gamma\right]$ is the propagation constant matrix, and $\left[{Z_c}\right]$ is the characteristic impedance matrix. Additionally, characteristic impedances can be defined as
\begin{subequations}
\begin{equation} \label{eq:Terminal_Zc_Equations}
\setlength{\arraycolsep}{1pt}
\left[ \begin{array}{cc}
{\vec{V}_T}^+ \\
{\vec{V}_T}^- \\
\end{array} \right] = 
\left[ \begin{array}{cc}
\left[{Z_c}_T\right] & \left[0\right]\\
\left[0\right] & -\left[{Z_c}_T\right]\\
\end{array} \right]
\left[ \begin{array}{cc}
{\vec{I}_T}^+ \\
{\vec{I}_T}^- \\
\end{array} \right]
\end{equation}
\begin{equation} \label{eq:Modal_Zc_Equations}
\setlength{\arraycolsep}{1pt}
\left[ \begin{array}{cc}
{\vec{V}_M}^+ \\
{\vec{V}_M}^- \\
\end{array} \right] = 
\left[ \begin{array}{cc}
\left[{Z_c}_M\right] & \left[0\right]\\
\left[0\right] & -\left[{Z_c}_M\right]\\
\end{array} \right]
\left[ \begin{array}{cc}
{\vec{I}_M}^+ \\
{\vec{I}_M}^- \\
\end{array} \right]
\end{equation}
\end{subequations}
Again applying equations (\ref{eq:Tv_Def}) and (\ref{eq:Ti_Def}), and comparing (\ref{eq:Terminal_Zc_Equations}) with (\ref{eq:Modal_Zc_Equations}) yields the relation
\begin{equation} \label{eq:Zc_Transform}
\left[{Z_c}_M\right] = \left[T_V\right]^{\text{-}1} \left[{Z_c}_T\right] \left[T_I\right]
\end{equation}
Other TL modal properties may be determined in a similar manner.

\subsection{Origins of Ambiguity}
\label{sec:Abiguity_Source}

Equations (\ref{eq:Z_Transform}) through (\ref{eq:Zc_Transform}) are well known and unambiguously correct. However, ambiguity is introduced in the solution process -- specifically, through the use of  (\ref{eq:G_Definition_1}) and (\ref{eq:G_Definition_2}), which are used to solve for the the transformation matrices $\left[T_V\right]$ and $\left[T_I\right]$ \emph{independent} of one another \cite{ref:Faria_Book}.
Specifically, since these two equations are of the form
\begin{equation} \label{eq:Example_Diagonalization}
\left[D\right] = \left[Q\right]^{\text{-}1} \left[M\right] \left[Q\right],
\end{equation}
where $\left[M\right]$ is a matrix to be diagonalized, $\left[Q\right]$ is the diagonalization matrix, and $\left[D\right]$ is a diagonal matrix said to be the diagonalized form of $\left[M\right]$, a regular eigenvalue process is typically used to simultaneously solve for $\left[Q\right]$ and $\left[D\right]$, given $\left[M\right]$. Unfortunately, the matrix $\left[Q\right]$ is not unique. It can be observed that if $\left[Q\right]$ is post-multiplied by a diagonal matrix $\left[g\right]$, such that
\begin{equation} \label{eq:Example_Diagonalization_SQg}
\left[S\right] = \left[Q\right] \left[g\right],
\end{equation}
then the diagonalization of $\left[M\right]$ by use of $\left[S\right]$ is
\begin{equation} \label{eq:Example_Diagonalization_S}
\left[D\right] = \left[S\right]^{\text{-}1} \left[M\right] \left[S\right],
\end{equation}
and is still valid, since
\begin{equation} \label{eq:Example_Diagonalization_Qg}
\left[D\right] = \left[g\right]^{\text{-}1} \left[Q\right]^{\text{-}1} \left[M\right] \left[Q\right] \left[g\right].
\end{equation}
A diagonal $\left[g\right]$ ensures the diagonality of $\left[D\right]$, since any $\left[Q\right]^{\text{-}1} \left[M\right] \left[Q\right]$ will itself be diagonal (subject to some physical constraints -- it has been shown that some physical systems can result in a non-diagonalizable matrix \cite{ref:Faria_NoEigenvalues}, although these are unlikely to be encountered in practice). Furthermore, it can be shown that since $\left[Q\right]^{\text{-}1} \left[M\right] \left[Q\right]$ is diagonal, the product $\left[g\right]^{\text{-}1} \left[Q\right]^{\text{-}1} \left[M\right] \left[Q\right] \left[g\right]$ reduces simply to $\left[Q\right]^{\text{-}1} \left[M\right] \left[Q\right]$, such that the value of $\left[g\right]$ has no bearing on the solution values $\left[D\right]$ and $\left[Q\right]$ in (\ref{eq:Example_Diagonalization_Qg}) \cite{ref:Faria_Book}.

This fact allows for the correct diagonalization of the propagation constants in (\ref{eq:G_Definition_1}) and (\ref{eq:G_Definition_2}), regardless of the values of $\left[g\right]$. The problem with this ambiguity arises from attempting to solve any of (\ref{eq:Z_Transform}), (\ref{eq:Y_Transform}), (\ref{eq:Zc_Transform}), for which $\left[T_V\right]$ and $\left[T_I\right]$ could each be post-multiplied by an arbitrary diagonal matrix -- however, in these equations, the value of $\left[g\right]$ will indeed have a direct impact on the diagonalized results, for example
\begin{equation} \label{eq:Zc_Transform2}
\left[{Z_c}_M\right] = \left[g_V\right]^{\text{-}1} \left[T_V\right]^{\text{-}1} \left[{Z_c}_T\right] \left[T_I\right] \left[g_I\right],
\end{equation}
the solution of which is presently ambiguous. In the following work the correction matrices $\left[g\right]$ will be utilized with the following definitions:
\begin{subequations}
\begin{equation} \label{eq:Tv_NewOld_Def}
\left[T_V\right]_{new} = \left[T_V\right]_{old} \left[g_V\right]
\end{equation}
\begin{equation} \label{eq:Ti_NewOld_Def}
\left[T_I\right]_{new} = \left[T_I\right]_{old} \left[g_I\right]
\end{equation}
\end{subequations}
where $\left[T_I\right]_{old}$ and $\left[T_V\right]_{old}$ are the matrices determined by the original eigenmode solution, and $\left[T_I\right]_{new}$ and $\left[T_V\right]_{new}$ are the corrected transformation matrices used to produce the correct modal TL properties. 

\subsection{Existing Disambiguation Processes}
\label{sec:Existing_Disambiguation_Processes}

Presently, there exist some strategies for solving the appropriate values of $\left[g_V\right]$ and $\left[g_I\right]$. Two of the most common are based on simple normalizations that are typically arbitrarily chosen and applied; both were detailed extensively in \cite{ref:Paul_Paper1}. The first is the normalization of the product $\left[T_V\right]^T \left[T_I\right]$, for which it was shown that the symmetric nature of $\left[\gamma_M\right]^2$, $\left[Z_T\right]$, $\left[Y_T\right]$ leads to the conclusion that 
\begin{equation} \label{eq:OldNormaization}
{\left[T_I\right]^T}_{new} {\left[T_V\right]}_{new} = {\left[T_V\right]^T}_{new} {\left[T_I\right]}_{new} = \left[D\right],
\end{equation}
where $\left[D\right]$ is any diagonal matrix, typically chosen to be identity for convenience. This is the introduction of one ambiguity, since the choice is entirely arbitrary. The second method involves a form of self-normalization of the matrix diagonals, such that 
\begin{equation} \label{eq:SelfNormaization}
{\left[T_I\right]^T}_{new} \left[T_I\right]_{new} = {\left[T_V\right]^T}_{new} \left[T_V\right]_{new} = \left[I\right]
\end{equation}
This process also introduces some ambiguity, since there is no physical justification for such a method. However, even though these processes are not rigorously justified or correct, they typically result in solutions with tolerable error, and therefore have been widely adopted. Some work has also been done in attempting to normalize via physical quantities such as voltage, current, and power, but this has generally resulted in non-unique solutions \cite{ref:Normalized_Eigenvalues1}.

\subsection{Initial Postulates and Terminology}
\label{sec:Postulates}
The definition of such a modal-terminal domain transformation relies on the fact that both domains are equally valid representations of the same physical system. Therefore, various physical properties must be equivalent in both domains, foremostly, total energy and total charge. In the frequency domain, these are directly related to total power and current. Accordingly, the following are postulated:
\newline 

\begin{itemize}
    \item That the total power being carried by a TL in the modal domain is equal to that carried in the terminal domain. This will be elaborated on in Sec. \ref{sec:PowerConservation}.
    \newline
    \item That the total co-directed currents in the modal domain are equal to the total of those in the terminal domain. This will be investigated in Sec. \ref{sec:CurrentConservation}.
    \newline

\end{itemize}

While the total voltages and currents in the terminal domain ($\vec{V}_T$ and $\vec{I}_T$, respectively) represent a superposition of excited modes, it will be of interest to examine the effects of a single excited mode in the terminal domain. In this case, these terminal voltages and currents will be expressed as:

\begin{subequations}
    \begin{equation} \label{eq:Terminal_Voltage_Mode_n}
        {\left.\vec{V}_T\right|}_n = \left[T_V\right] \vec{\delta}_n {V_M}_n
    \end{equation}
        \begin{equation} \label{eq:Terminal_Current_Mode_n}
        {\left.\vec{I}_T\right|}_n = \left[T_I\right] \vec{\delta}_n {I_M}_n
    \end{equation}
\end{subequations}
where ${V_M}_n$ and ${I_M}_n$ are the voltage and current of the excited mode $n$, respectively, and the delta vector $\vec{\delta}_n$ is defined as
\begin{equation} \label{eq:Delta_Definition}
    \vec{\delta}_n = 
    \begin{cases}
        1, & \text{if } i = n \\
        0, & \text{otherwise} \\
    \end{cases}
    \forall~\text{indices}~i.
\end{equation}
The complex-conjugate transpose of a matrix or vector will be expressed as
\begin{equation} \label{eq:ConjugateTranspose}
\left[A\right]^* = \overline{\left[A\right]}^T,
\end{equation}
where the over-bar denotes the element-wise complex conjugate. The inner product will be used to represent sums over vectors; that is, for vectors $\vec{a}$ and $\vec{b}$
\begin{equation} \label{eq:Sum_InnerProduct_Definition}
    \sum_k{a_k b_k} = \vec{a} \cdot \vec{b} = \vec{a}^T \vec{b}.
\end{equation}
In this manner, the sum of the elements in a single vector may be expressed as
\begin{equation} \label{eq:Sum_Definition}
    \sum_k{a_k} = \vec{1}^T \vec{a},
\end{equation}
where $\vec{1}$ is a vector, the entries of which are each unity.

\subsection{Power Equivalence}
\label{sec:PowerConservation}

In accordance with the first postulate,
\begin{equation} \label{eq:Modal_Terminal_Power}
    \vec{I}_M^* \vec{V}_M  = \vec{I}_T^* \vec{V}_T
\end{equation}
Substituting the terminal to modal domain transforms in (\ref{eq:Tv_Def}) and (\ref{eq:Ti_Def}) yields
\begin{equation} \label{eq:Modal_Terminal_Power_Equated}
    \vec{I}_M^* \vec{V}_M = \vec{I}_M^* \left[T_I\right]_{new}^* \left[T_V\right]_{new} \vec{V}_M 
\end{equation}
to which one of the solutions is the identity matrix $\left[I\right]$. Furthermore, equating the carried power by a single mode to its terminal-domain equivalent yields
\begin{equation} \label{eq:Mode_Terminal_Power_Equated}
    {I_M^*}_n {V_M}_n = {\left.\vec{I}_T^*\right|}_n {\left.\vec{V}_T\right|}_n = {I_M^*}_n \vec{\delta}_n^T \left[T_I\right]_{new}^* \left[T_V\right]_{new} \vec{\delta}_n {V_M}_n,
\end{equation}
from which the scalars can be cancelled to give
\begin{equation} \label{eq:Mode_Terminal_Power_Equated2}
    \vec{\delta}_n^T \left[T_I\right]_{new}^* \left[T_V\right]_{new} \vec{\delta}_n  = 1,
\end{equation}
which, in conjunction with (\ref{eq:Modal_Terminal_Power_Equated}) demonstrates the unique definition of 
\begin{equation} \label{eq:Modal_Terminal_Power_Condition}
    \left[T_I\right]_{new}^* \left[T_V\right]_{new} = \left[I\right]
\end{equation}

A similar derivation was investigated in \cite{ref:Paul_Paper1}, however, this was only for one component of power (the real part), and subsequently an equation such as (\ref{eq:Mode_Terminal_Power_Equated2}) could not be developed through the cancellation of complex voltage and current terms to demonstrate the identity matrix as the unique solution to (\ref{eq:Modal_Terminal_Power_Condition}).

\subsection{Current Equivalence}
\label{sec:CurrentConservation}

Consider first a single excited mode, $n$, which is modelled via a two-conductor TL. The current magnitude for this mode is expressed, in accordance with the previous definitions, as ${I_M}_n$. While this is a single quantity, there are of course two currents present: one on each conductor. Let the magnitude of the current on one of the conductors be labelled ${I_M}_n\left(+\right)$, and the magnitude of the current flowing in the opposite direction -- that is, on the other conductor -- be labelled ${I_M}_n\left(-\right)$. The conservation of current enforces the equality of the currents ${I_M}_n\left(+\right) = {I_M}_n\left(-\right)$.

Consider next a TL with more than two conductors, as described in the terminal domain. Assume for now that the current carried by the reference conductor is included in the set of all currents. It is certain that the conservation of current specifies that the sum of the currents on all conductors must be zero. Similar to the modal domain TL, assume that the current magnitudes may be divided into two sets $\vec{I}_T\left(+\right)$ and $\vec{I}_T\left(-\right)$, with the membership of the sets being determined by the phase of these currents with respect to the modal currents. Although it may seem intuitive for TEM modes, a proof that only two such contra-directed phases (corresponding to co- and contra-directed currents) exist is supplied in Appendix \ref{sec:AppendixContraDirectedPhases}. According to the conservation of current, the sums of each group must be equal to each other, that is: $\vec{1}^T \vec{I}_T\left(+\right) = \vec{1}^T \vec{I}_T\left(-\right)$. Furthermore, the set of total currents on all conductors is equal to the difference of the two groups, such that
\begin{equation} \label{eq:Current_Equivalences_Sum}
    \vec{I}_T = \vec{I}_T\left(+\right) - \vec{I}_T\left(-\right)
\end{equation}
The vector $\vec{I}_T\left(+\right)$ will then contain an entry of zero wherever $\vec{I}_T\left(-\right)$ has a nonzero entry, and vice-versa, such that all vectors have the same length (which is the number of conductors in the TL). 

It is known from (\ref{eq:Ti_Def}) that the distribution of terminal-domain currents is dictated by the transformation matrix $\left[T_I\right]$ -- that is, if a single mode is excited, the currents on each conductor are uniquely specified by the corresponding column of $\left[T_I\right]$. Let the transformation matrix then be separated into two new matrices -- one, labelled $\left[T_I(+)\right]$, consisting of the components related to the set of currents $\vec{I}_T\left(+\right)$ and another labelled $\left[T_I(-)\right]$, consisting of the components related to the set of currents $\vec{I}_T\left(-\right)$. Then,

\begin{subequations}
    \begin{equation} \label{eq:Current_Equivalences_Sum}
        \left[T_I\right] = \left[T_I(+)\right] - \left[T_I(-)\right]
    \end{equation}
    \begin{equation} \label{eq:Current_Equivalences_Positive}
        \vec{I}_T\left(+\right) = \left[T_I(+)\right] \vec{I}_M
    \end{equation}
    \begin{equation} \label{eq:Current_Equivalences_Negative}
        \vec{I}_T\left(-\right) = \left[T_I(-)\right] \vec{I}_M
    \end{equation}
\end{subequations}
The equality of currents between domains specifies that the total co-directed currents must be equal in both the modal and terminal domains. Specifically, for an excited mode $n$, there must exist appropriate sets $\vec{I}_T\left(+\right)_{new}$ and $\vec{I}_T\left(-\right)_{new}$, such that

\begin{subequations}
    \begin{equation} \label{eq:Conservation_of_Charge_Positive}
        {I_M}_n\left(+\right) = \vec{1}^T {\left.\vec{I}_T\left(+\right)_{new}\right|}_n
    \end{equation}
    \begin{equation} \label{eq:Conservation_of_Charge_Negative}
        {I_M}_n\left(-\right) = \vec{1}^T {\left.\vec{I}_T\left(-\right)_{new}\right|}_n
    \end{equation}
\end{subequations}
Invoking (\ref{eq:Terminal_Current_Mode_n}) and (\ref{eq:Current_Equivalences_Positive}) allows the previous expressions to be expanded to

\begin{subequations}
    \begin{equation} \label{eq:Conservation_of_Charge_Positive2}
        {I_M}_n\left(+\right) = \vec{1}^T \left[T_I(+)\right]_{new} \vec{\delta}_n {I_M}_n\left(+\right)
    \end{equation}
    \begin{equation} \label{eq:Conservation_of_Charge_Negative2}
        {I_M}_n\left(-\right) = \vec{1}^T \left[T_I(-)\right]_{new} \vec{\delta}_n {I_M}_n\left(-\right)
    \end{equation}
\end{subequations}
The scalars ${I_M}_n\left(+\right)$ and ${I_M}_n\left(-\right)$ can then be cancelled to yield the expressions:

\begin{subequations}
    \begin{equation} \label{eq:Conservation_of_Charge_Positive3}
        1 = \vec{1}^T \left[T_I(+)\right]_{new} \vec{\delta}_n
    \end{equation}
    \begin{equation} \label{eq:Conservation_of_Charge_Negative3}
        1 = \vec{1}^T \left[T_I(-)\right]_{new} \vec{\delta}_n
    \end{equation}
\end{subequations}
which simply imply that the sum of each column in $\left[T_I(+)\right]_{new}$ and $\left[T_I(-)\right]_{new}$ must be equal to unity.

The unique correction process may then be determined by substituting (\ref{eq:Current_Equivalences_Sum}) into (\ref{eq:Ti_NewOld_Def}), and letting the matrices $\left[T_I(+)\right]_{old}$ and $\left[T_I(-)\right]_{old}$ be specified to contain the same non-zero indices as $\left[T_I(+)\right]_{new}$ and $\left[T_I(-)\right]_{new}$,

\begin{subequations}
\begin{equation} \label{eq:T_I_Positive_NewOld_Def}
\left[T_I(+)\right]_{new} = \left[T_I(+)\right]_{old} \left[g_I\right]
\end{equation}
\begin{equation} \label{eq:T_I_Negative_NewOld_Def}
\left[T_I(-)\right]_{new} = \left[T_I(-)\right]_{old} \left[g_I\right]
\end{equation}
\end{subequations}
Substituting the previous equations into (\ref{eq:Conservation_of_Charge_Positive3}) and (\ref{eq:Conservation_of_Charge_Negative3}) yields:

\begin{subequations}
    \begin{equation} \label{eq:Conservation_of_Charge_Positive_Old}
        1 = \vec{1}^T \left[T_I(+)\right]_{old} \left[g_I\right] \vec{\delta}_n
    \end{equation}
    \begin{equation} \label{eq:Conservation_of_Charge_Negative_Old}
        1 = \vec{1}^T \left[T_I(-)\right]_{old} \left[g_I\right] \vec{\delta}_n
    \end{equation}
\end{subequations}
which indicates that the entries along the diagonal of the correction matrix $\left[g_I\right]$ are simply the inverse of the sum of the corresponding columns in $\left[T_I(+)\right]_{old}$ and $\left[T_I(-)\right]_{old}$.

Recall that in practical use, the information related to the reference conductor would not be contained directly in $\vec{I}_{Tnew}$, $\left[T_I\right]_{new}$, $\left[T_I(+)\right]_{new}$, or $\left[T_I(-)\right]_{new}$. This is not such a critical issue, as being a single, scalar value, it would, if accounted for, either appear in the set of currents $\vec{I}_T\left(+\right)_{new}$ or $\vec{I}_T\left(-\right)_{new}$, and subsequently be manifested in either of $\left[T_I(+)\right]_{old}$ or $\left[T_I(-)\right]_{old}$. This does mean that, in practice, only one of (\ref{eq:Conservation_of_Charge_Positive_Old}) or (\ref{eq:Conservation_of_Charge_Negative_Old}) will be correct -- but the issue is simply resolved by noting that the correct equation will always correspond to the sum with the larger magnitude. Then,

\begin{equation} \label{eq:Conservation_of_Charge_Correction}
    g_{In} = max\left[\left(\vec{1}^T \left[T_I(+)\right]_{old} \vec{\delta}_n\right), \left(\vec{1}^T \left[T_I(-)\right]_{old} \vec{\delta}_n\right)\right]^{-1}
\end{equation}
and hence the current transformation matrix is uniquely defined.

It is worth noting that if the transformation matrices (in their corrected state) are complex, several contradictions may arise with the previously derived equations. To alleviate such concerns, Appendix \ref{sec:AppendixRealMatrices} offers a proof of the remarkable fact that the transformation matrices must be real -- indicating that the terminal-domain voltages and currents must always be in phase with their corresponding modal quantities.

\subsection{Final Solution Process}
\label{sec:FinalSolutionProcess}

The solution process to determine the corrected, unique transformation matrices is then as follows:

\begin{enumerate}

    \item Find the original, uncorrected, transformation matrices $\left[T_I\right]_{old}$ and $\left[T_V\right]_{old}$  (for example, via an eigenmode process utilizing (\ref{eq:G_Definition_1}) and/or (\ref{eq:G_Definition_2})).
    
    \item Apply the equality of currents to obtain $\left[g_I\right]$ from $\left[T_I\right]_{old}$, utilizing (\ref{eq:Conservation_of_Charge_Correction}). 

    \item Apply (\ref{eq:Ti_NewOld_Def}) to obtain the correct current transformation matrix $\left[T_I\right]_{new}$ from $\left[g_I\right]$ and $\left[T_I\right]_{old}$.

    \item Employ the equality of power (\ref{eq:Modal_Terminal_Power_Condition}) to obtain the corrected voltage transformation matrix $\left[T_V\right]_{new}$, that is, $\left[T_V\right]_{new} = \left[T_I\right]_{new}^{-1*}$.

\end{enumerate}
A proposed algorithm for correcting the transformation matrices and associated discussion are given in Appendix \ref{sec:AppendixAlgorithm}.

\section{Examples}
\label{sec:Examples}

This section will examine four examples of TLs in the TEM approximation, and perform the process used to diagonalize their properties, as explained in the previous section. Ansys HFSS was used to extract the per-unit-length inductance and capacitance matrices in the terminal domain, as well as determine the propagation constants and characteristic impedances in the modal domain, all computed at a frequency of 1~GHz. The convergence criteria for the HFSS simulations were chosen to be extremely strict, since the produced data will be the only standard of comparison in this work. These criteria enforced a minimum of 20 converged passes with a convergence $\Delta\left|S\right|$ of 0.0001 and $\Delta\angle S$ of 0.1$^\circ$. The final data may still have errors of roughly 1$\%$, due to extensive numerical processing. Data which are presented in the form of complex inductance ([\emph{L}]) and capacitance ([\emph{C}]) matrices include losses from the traditional resistance ([\emph{R}]) and conductance ([\emph{G}]). The validity and effectiveness of the proposed process is established by the corroboration of the HFSS modal values, given only the known terminal values.

\subsection{Three-Wire Line}
\label{sec:Three-Wire Line}

\begin{figure}
\centering
\includegraphics[width=2.4in]{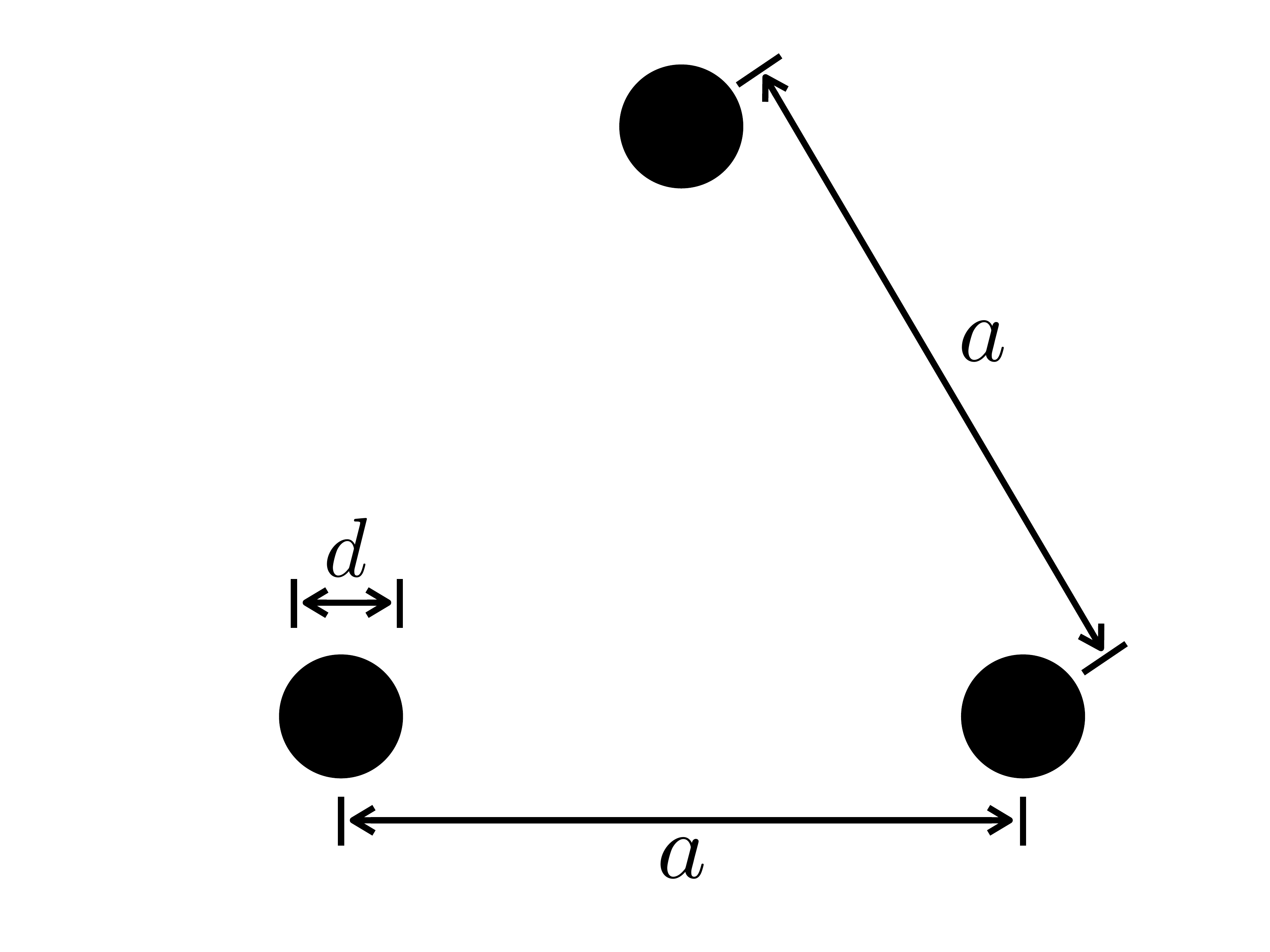}
\caption{Setup of the three wire transmission line. The conductors (black circles) are similar and form an equilateral triangle.}
\label{fig:ThreeWireLine_Setup}
\end{figure}

This TL consists of three cylindrical copper conductors of diameter $d$~=~1.00 mm suspended in air and spaced equally apart by a distance $a$~=~3.46 mm, as shown in Fig.~\ref{fig:ThreeWireLine_Setup}. Since the system is symmetric, the choice of reference conductor is arbitrary, and the $\left[L_T\right]$ and $\left[C_T\right]$ matrices are 2$\times$2 in size, with
\begin{equation} \label{eq:Example1_LtCt}
\begin{gathered}
\left[ L_T \right] = 
\left[ \begin{array}{cc}
0.7611 - j0.0009 & 0.3799 - j0.0004 \\
0.3799 - j0.0004 & 0.7611 - j0.0009 \\
\end{array} \right] uH/m
\\
\left[ C_T \right] = 
\left[ \begin{array}{cc}
19.4946 - j0.0002 & - 9.7654 + j0.0002 \\
- 9.7654 + j0.0002 & 19.4946 - j0.0003 \\
\end{array} \right] pF/m
\end{gathered}
\end{equation}
Employing equations (\ref{eq:G_Definition_1}) and (\ref{eq:G_Definition_2}) (noting that these two equations are solved \emph{independently} of one another) yields the modal propagation constants $\left[\gamma_M\right]$, as well as the current and voltage transformation matrices:
\begin{equation} \label{eq:Example1_GmTiTv}
\begin{gathered}
\left[ \gamma_M \right] = 
\left[ \begin{array}{cc}
0.0130 + j20.9343 & 0.0000 + j0.0000 \\
0.0000 + j0.0000 & 0.0122 + j20.9864 \\
\end{array} \right] /m
\\
\left[ T_V \right]_{old} = 
\left[ \begin{array}{cc}
0.7071 + j0.0094 & -0.7071 - j0.0000 \\
0.7071 + j0.0098 &  0.7071 - j0.0012 \\
\end{array} \right]
\\
\left[ T_I \right]_{old} = 
\left[ \begin{array}{cc}
0.7071 + j0.0094 & -0.7071 - j0.0000 \\
0.7071 + j0.0105 &  0.7071 - j0.0004 \\
\end{array} \right]
\end{gathered}
\end{equation}
where it can be noted that the uncorrected transformation matrices within acceptable error satisfy the equality of total power (\ref{eq:Modal_Terminal_Power_Condition}). The modal propagation constants agree to within 0.2\% with the modal solutions given by HFSS, which are
\begin{equation} \label{eq:Example1_HFSSGm}
\left[ \gamma_M \right] = 
\left[ \begin{array}{cc}
0.0122 + j20.9707 & 0.0000 + j0.0000 \\
0.0000 + j0.0000 & 0.0122 + j20.9707 \\
\end{array} \right] /m
\end{equation}
Subsequently, the terminal and (incorrect) modal characteristic impedances, the latter determined from either (\ref{eq:Zc_Definition_1}) or (\ref{eq:Zc_Definition_2}), are
\begin{equation} \label{eq:Example1_ZctZcm}
\begin{gathered}
\left[ Z_{cT} \right] = 
\left[ \begin{array}{cc}
228.3030 - j0.1324 & 114.1510 - j0.0662 \\
114.1510 - j0.0662 & 228.3030 - j0.1324 \\
\end{array} \right] \Omega
\\
\left[ Z_{cM} \right] = 
\left[ \begin{array}{cc}
342.4540 - j0.0083 & 0.0000 - j0.0000 \\
0.0000 - j0.0000 & 114.1520 - j0.0027 \\
\end{array} \right] \Omega
\end{gathered}
\end{equation}
the latter of which are significantly different from the HFSS modal characteristic impedance values of
\begin{equation} \label{eq:Example1_HFSSZcm}
\left[ Z_{cM} \right] = 
\left[ \begin{array}{cc}
228.9177 - j0.1178 & 0.0000 - j0.0000 \\
0.0000 - j0.0000 & 171.6887 - j0.0884 \\
\end{array} \right] \Omega
\end{equation}
exhibiting differences of 39.7\% and 40.3\%, respectively. This is expected, as the current and voltage transformation matrices $\left[T_I\right]$ and $\left[T_V\right]$ do not strictly respect current or power equivalence between domains, for which they must be first corrected using the proposed processes.

The correction process is then implemented by utilizing (\ref{eq:Tv_NewOld_Def}), (\ref{eq:Ti_NewOld_Def}), (\ref{eq:Modal_Terminal_Power_Condition}), and (\ref{eq:Conservation_of_Charge_Correction}). Specifically, implementing the latter gives
\begin{equation} \label{eq:Example1_gI}
\left[ g_I \right] = 
\left[ \begin{array}{cc}
0.7071 - j0.0099 & 0.0000 + j0.0000 \\
0.0000 + j0.0000 & 1.4142 + j0.0008  \\
\end{array} \right]
\end{equation}
Then, it follows that the corrected transformation matrices are 
\begin{equation} \label{eq:Example1_TiTvNew}
\begin{gathered}
\left[ T_V \right]_{new} = 
\left[ \begin{array}{cc}
1.0000 + j0.0003 & -0.5000 + j0.0006 \\
1.0000 - j0.0003 &  0.5000 - j0.0003 \\
\end{array} \right]
\\
\left[ T_I \right]_{new} = 
\left[ \begin{array}{cc}
0.5000 - j0.0004 & -1.0000 + j0.0000 \\
0.5000 + j0.0004 &  1.0000 - j0.0005 \\
\end{array} \right]
\end{gathered}
\end{equation}
which are well within the acceptable error of being entirely real. Using these new values in (\ref{eq:Zc_Transform}) gives
\begin{equation} \label{eq:Example1_Zcm_Corrected}
\left[ Z_{cM} \right] = 
\left[ \begin{array}{cc}
228.3020 - j0.2593 & 0.000 + j0.1906 \\
-0.0001 - j0.1902 & 171.2270 - j0.0993 \\
\end{array} \right] \Omega
\end{equation}
which are in much closer agreement (a percent difference of 0.27\% for both values on the diagonal) to the modal values given by HFSS (\ref{eq:Example1_HFSSZcm}) than the uncorrected values in (\ref{eq:Example1_ZctZcm}). For additional verification, it can be shown that, within the given error, $\left[ T_V \right]_{new}^T \left[ T_I \right]_{new}^* = \left[I \right]$, and that $\left[ \gamma_M \right]$ remains consistent whether  solved with (\ref{eq:G_Definition_1}) or (\ref{eq:G_Definition_2}).

The current and voltage transformation matrices contain information regarding the relative voltage and current distribution of each mode in each column. Observing the first column (mode), it is noted in $\left[ T_I \right]$ that the relative currents have equal magnitudes and are in the same direction. By deduction, it may be inferred that the reference conductor then carries a relative  current with double the magnitude of either of the other two conductors, and from $\left[ T_V \right]$ it is observed that the two non-reference conductors have the same relative voltage. This mode then corresponds to a ``common" mode, where the two-non reference conductors have the same relative  current and voltage characteristics. Observing the second mode, it can be seen that in the corresponding second column of $\left[ T_I \right]$, there are opposite and equal-magnitude relative currents on the non-reference conductors, leading to the conclusion that the relative current on the reference conductor is zero. Since the second column entries in $\left[ T_V \right]$ are also equal in magnitude, but oppositely directed, it can be inferred that this is a ``balanced" (or ``differential") mode.

\subsection{Shielded, Conductor-Backed, Coplanar Waveguide}
\label{sec:S-CBCPW}

\begin{figure}
\centering
\includegraphics[width=3.0in]{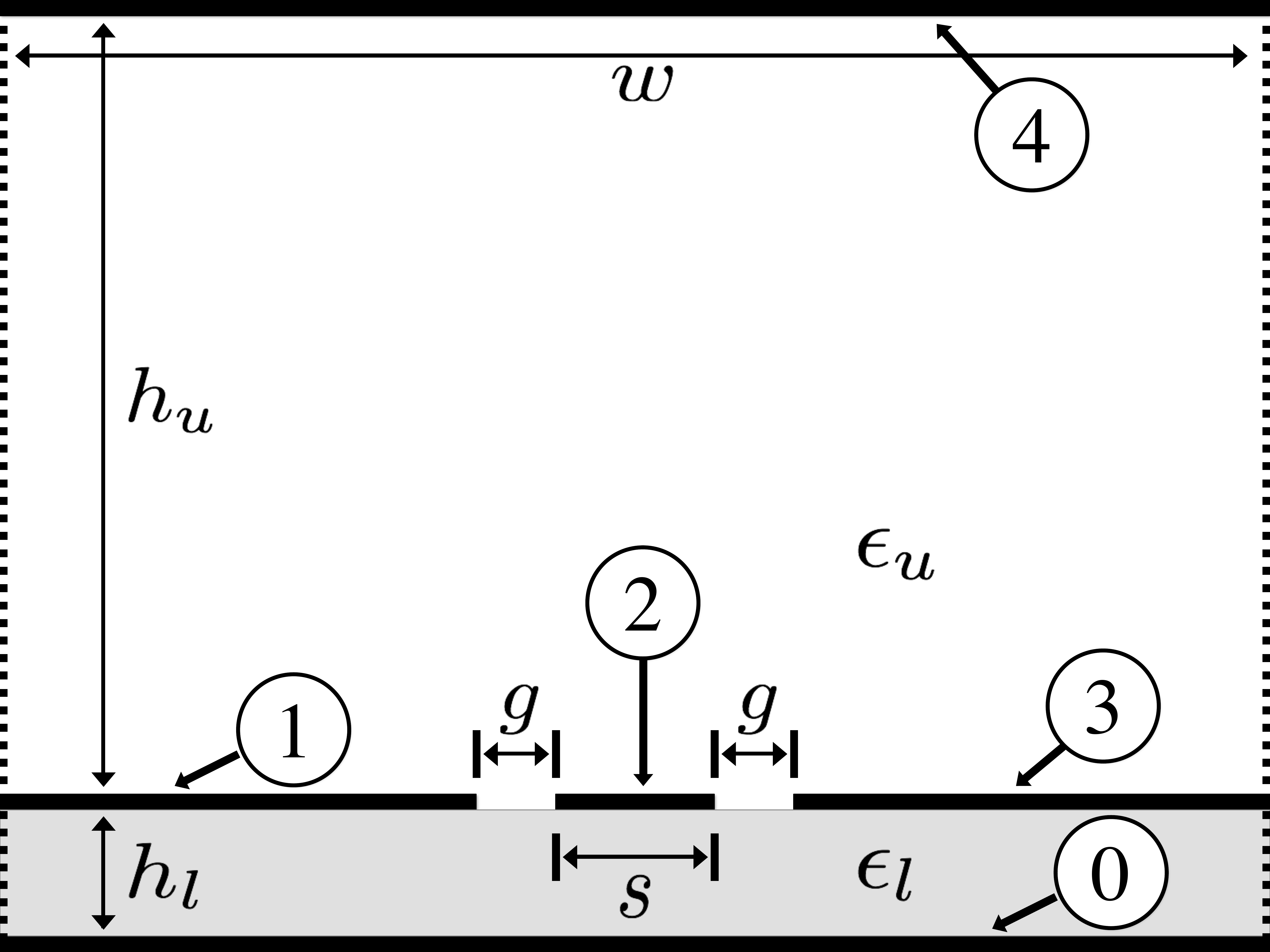}
\caption{Setup of the S-CBCPW transmission line. The conductors (solid black rectangles) are numbered, with the conductor backing (\#1) taken as the reference.}
\label{fig:SCBCPW_Setup}
\end{figure}

The shielded, conductor-backed coplanar waveguide (S-CBCPW) TL used is shown in Fig. \ref{fig:SCBCPW_Setup}. The lower dielectric is RO-3010, and the upper dielectric is air. The boundary conditions used on the vertical sides are (fictitious) perfect magnetic conductors (PMCs), indicated by the dashed lines. The numbers indicate the conductors, with number 0 (the conductor backing) being taken as the reference. Conductors 1 and 3 are the CPW grounds, while conductor 2 is the CPW strip line, and conductor 4 is the shield. The various parameters used were $h_l$ = 1.524mm, $h_u$ = 100mm, $s$ = 1.00mm, $g$ = 1.00mm, $w$ = 10mm, $\epsilon_l$ $\approx$ 10.2, $\epsilon_u$ $\approx$ 1.00. The imaginary components of all matrices will be ignored for this example, as they are negligibly small. Simulation in HFSS yields the matrices with real components
\begin{equation} \label{eq:Example2_LtCt}
\begin{gathered}
\left[ L_T \right] = 
\left[ \begin{array}{cccc}
0.3613 & 0.1269 & 0.0640 & 0.1928 \\
0.1269 & 0.5000 & 0.1269 & 0.1925 \\
0.0640 & 0.1269 & 0.3613 & 0.1928 \\
0.1928 & 0.1925 & 0.1928 & 12.7667 \\
\end{array} \right] \mu H/m
\\
\left[ C_T \right] = 
\left[ \begin{array}{cccc}
262.1345 & -26.5954 & - 4.2907 & - 0.3507 \\
-26.5954 & 157.3421 & -26.5954 & - 0.1664 \\
 -4.2904 & -26.5954 & 262.1345 & - 0.3507\\
- 0.3507 & - 0.1664 & - 0.3507 & 0.8856 \\
\end{array} \right] pF/m
\end{gathered}
\end{equation}
These yield modal propagation constants and (uncorrected) transformation matrices
\begin{equation} \label{eq:Example2_Gm}
\left[ \gamma_M \right] = j
\left[ \begin{array}{cccc}
66.2749 & 0.0000  & 0.0000  & 0.0000 \\
0.0000  & 55.9251 & 0.0000  & 0.0000 \\
0.0000  & 0.0000  & 50.0294 & 0.0000 \\
0.0000  & 0.0000  & 0.0000  & 20.9404\\
\end{array} \right] /m
\end{equation}
and transformation matrices
\begin{equation} \label{eq:Example2_TiTv}
\begin{gathered}
\left[ T_I \right]_{old} = 
\left[ \begin{array}{cccc}
0.6755 &  0.7071 &-0.4036 &-0.3435 \\
0.2959 &  0.0000 & 0.8210 &-0.1559 \\
0.6755 & -0.7071 &-0.4036 &-0.3435 \\
0.0000 &  0.0000 & 0.0000 & 0.8600 \\
\end{array} \right]
\\
\left[ T_V \right]_{old} = 
\left[ \begin{array}{cccc}
0.5000 &-0.7071 &-0.2092 & 0.0000 \\
0.5000 & 0.0000 & 0.9552 & 0.0000 \\
0.5000 & 0.7071 &-0.2092 & 0.0000 \\
0.5000 & 0.0000 & 0.0060 & 1.0000 \\
\end{array} \right]
\end{gathered}
\end{equation}
The modal propagation constants are once again very close (within 0.2\%) to the HFSS-given values, which have imaginary components of
\begin{equation} \label{eq:Example2_HFSSGm}
\left[ \gamma_M \right] = j
\left[ \begin{array}{cccc}
66.2770 & 0.0000  & 0.0000  & 0.0000 \\
0.0000  & 55.9230 & 0.0000  & 0.0000 \\
0.0000  & 0.0000  & 50.0230 & 0.0000 \\
0.0000  & 0.0000  & 0.0000  & 20.9680\\
\end{array} \right] /m
\end{equation}
The (real components of the incorrect) modal characteristic impedances derived using the above transformation matrices are
\begin{equation} \label{eq:Example2_Zcm}
\left[ Z_{cM} \right] = 
\left[ \begin{array}{cccc}
33.4080 & 0.0000 & 0.0000 & 0.0000 \\
0.0000 & 61.4095 & 0.0000 & 0.0000 \\
0.0000 & 0.0000 & 40.5009 & 0.0000 \\
0.0000 & 0.0000 & 0.0000 & 3241.0304 \\
\end{array} \right] \Omega
\end{equation}
which are not very similar to those given by HFSS:
\begin{equation} \label{eq:Example2_HFSSZcm}
\left[ Z_{cM} \right] = 
\left[ \begin{array}{cccc}
18.6530 & 0.0000 & 0.0000 & 0.0000 \\
0.0000 & 66.9290 & 0.0000 & 0.0000 \\
0.0000 & 0.0000 & 58.6080 & 0.0000 \\
0.0000 & 0.0000 & 0.0000 & 3842.9500 \\
\end{array} \right] \Omega
\end{equation}
with percentage differences of 8.6\%, 56.7\%, 36.5\%, and 17.0\% respectively. The correction process yields the matrices
\begin{equation} \label{eq:Example2_gTiTvNew}
\begin{gathered}
\left[ g_I \right] = 
\left[ \begin{array}{cccc}
0.6072 & 0.0000 & 0.0000 & 0.0000 \\
0.0000 & 1.4142 & 0.0000 & 0.0000 \\
0.0000 & 0.0000 & 1.2180 & 0.0000 \\
0.0000 & 0.0000 & 0.0000 & 1.1628 \\
\end{array} \right]
\\
\left[ T_I \right]_{new} = 
\left[ \begin{array}{cccc}
0.4102 &  1.0000 &-0.4916 &-0.3994 \\
0.1797 &  0.0000 & 1.0000 &-0.1813 \\
0.4102 & -1.0000 &-0.4916 &-0.3994 \\
0.0000 &  0.0000 & 0.0000 & 1.0000 \\
\end{array} \right]
\\
\left[ T_V \right]_{new} = 
\left[ \begin{array}{cccc}
1.0000 &  0.5000 &-0.1860 & 0.0000 \\
1.0000 &  0.0000 & 0.8086 & 0.0000 \\
1.0000 & -0.5000 &-0.1860 & 0.0000 \\
1.0000 &  0.0000 &-0.0021 & 1.0000 \\
\end{array} \right]
\end{gathered}
\end{equation}
which, in turn, allow for the computation of the corrected characteristic impedance matrix, the real values of which are
\begin{equation} \label{eq:Example2_Zcm_Corrected}
\left[ Z_{cM} \right] = 
\left[ \begin{array}{cccc}
18.6410 & 0.0000  & 0.0000  & 0.0000   \\
0.0000  & 66.8159 & 0.0000  & 0.0000   \\
0.0000  & 0.0000  & 58.2410 & 0.0000   \\
0.0000  & 0.0000  & 0.0000  & 3844.5347\\
\end{array} \right] \Omega
\end{equation}
the values of which are very close to those given by HFSS in (\ref{eq:Example2_HFSSZcm}): percentage differences of 0.17\%, 0.06\%, 0.63\%, and 0.04\% respectively.

The nature of the various modes can be determined from observation of the transformation matrices. The first mode (column) is characterized by a relative current value of approximately 0.4 on each of the CPW ground conductors (numbers 1 and 3), and approximately 0.2 on the CPW strip line (conductor 2). There is a negligibly small relative current component on the shield (conductor 4), and the same relative voltage is present on conductors 1 through 4, indicating that there is a negligibly small electric field in the upper region (the region between the plane of conductors 1, 2 and 3, and the plane of conductor 4). Since the same relative voltage exists on the three CPW conductors (1 through 3), it can be concluded that this mode is a parallel-plate-waveguide (PPW)-type mode in the lower dielectric region. The relative current magnitudes are a result of a nearly equally distributed (relative) current density over unequally sized conductors.

The second mode can be analyzed simply, as only two of the conductors -- the two CPW ground planes (conductors 1 and 3) -- support non-zero relative currents and voltages. This corresponds to the coupled-slotline (CSL) mode, for which there are indeed currents induced on the remaining conductors, but the net currents on each conductor are zero.

The third mode is characterized by a positive relative current on the CPW strip line (conductor 2), and negative relative currents of nearly half that magnitude on each of the CPW grounds (conductors 1 and 3). This is what would be expected of a CPW mode, although it can be observed that the sum of the negative relative currents implies that there is some small relative current component on the reference conductor backing as well. This would be expected from a conductor-backed CPW mode, and also it can be observed from the relative voltages that there is some coupling with the shield as well.

The last mode is similar to the first, in that all of the conductors except for one are equipotential. This, along with the relative currents on all conductors, readily identifies this mode as the PPW-like mode in the upper region.

\subsection{Lossy Two-Layer Parallel-Plate Waveguide}
\label{sec:2LayerPPW}

\begin{figure}
\centering
\includegraphics[width=3.0in]{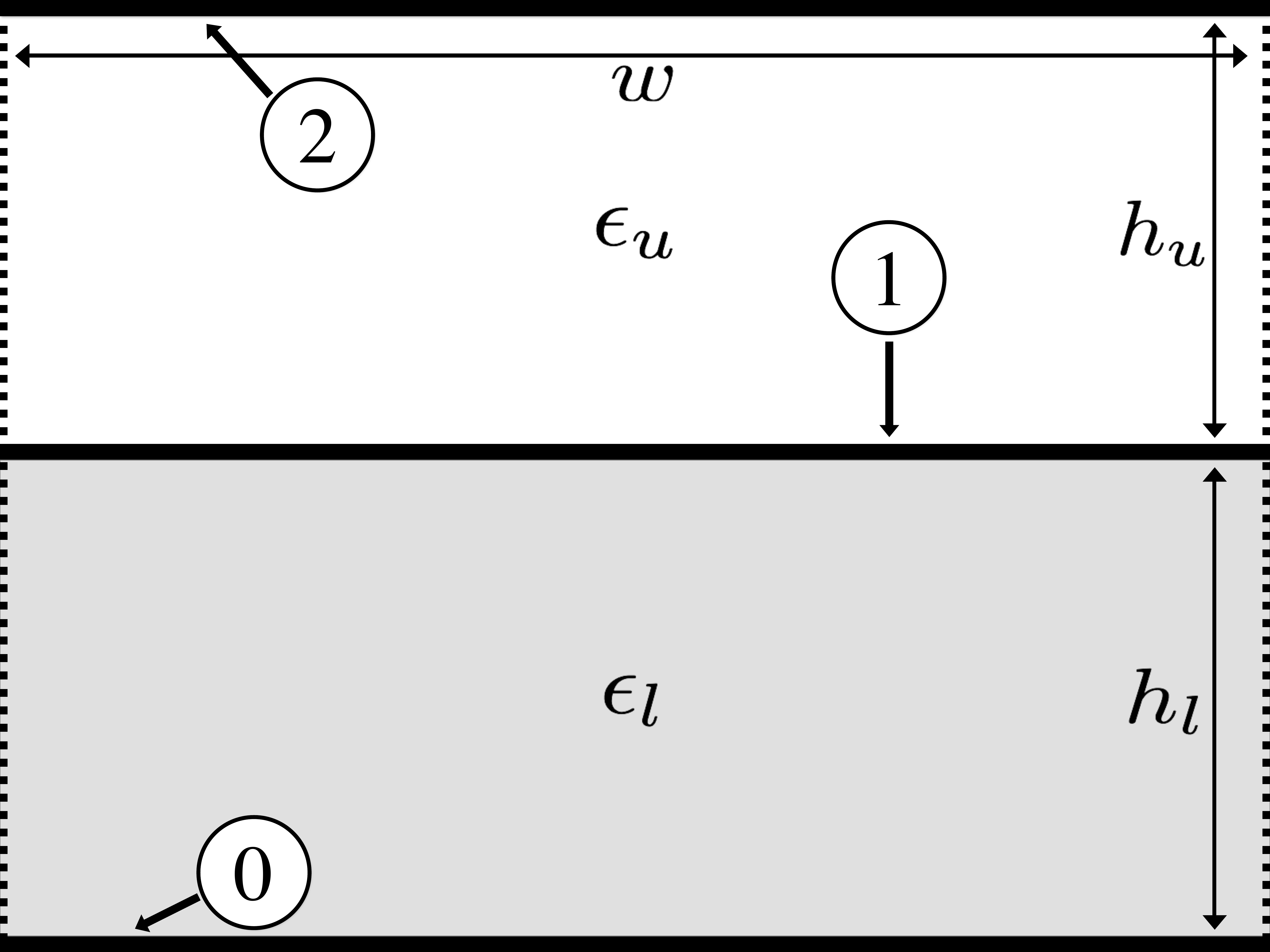}
\caption{Setup of the two-layer PPW transmission line. The conductors (solid black rectangles) are numbered, with the conductor backing (conductor 0) taken as the reference.}
\label{fig:TwoLayerPPW_Setup}
\end{figure}

To confirm the validity of the proposed solution, this example will deal with artificially inflated losses. The TL consists of three rectangular conductors, as shown in Fig. \ref{fig:TwoLayerPPW_Setup}, which are assumed to be perfectly conducting. It is also assumed that the upper and lower faces of the middle conductor (conductor 1) possess the same voltage and current values, such that there is coupling between the upper and lower regions. The vertical boundary conditions are again modelled as PMCs. Let $w$ = 10 mm, $h_l$ = 1.524 mm, $h_u$ = 10 mm, $\epsilon_l$ = 3.0 $-\ j$3.0, and $\epsilon_u$ = 1.0 $-\ j$1.0. Under these conditions, the terminal domain parameters are solved as

\begin{equation} \label{eq:Example3_LtCt}
\begin{gathered}
\left[L_T \right] = 
\left[\begin{array}{cc}
0.1918 - j0.0003 & 0.1918 - j0.0003 \\
0.1918 - j0.0003 & 1.4487 - j0.0003 \\
\end{array} \right] \mu H/m
\\
\left[C_T \right] = 
\left[\begin{array}{cc}
183.1490 - j183.1490 & -8.8542 + j8.8542 \\
-8.8542 + j8.8542 & 8.8542 - j8.8542 \\
\end{array} \right] pF/m
\end{gathered}
\end{equation}
These data give modal propagation-constant and (uncorrected) modal characteristic-impedance and transformation matrices of
\begin{equation} \label{eq:Example3_UncorrectedGmZcmTiTv}
\begin{gathered}
\left[ \gamma_M \right] = 
\left[ \begin{array}{cc}
9.5390 + j23.0291 & 0.0000 + j0.0000 \\
0.0000 + j0.0000 & 16.5589 + j39.8994 \\
\end{array} \right] /m
\\
\left[ T_I \right]_{old} = 
\left[ \begin{array}{cc}
-0.7071 + j0.0000 & 1.0000 + j0.0000 \\
0.7071 + j0.0000 & 0.0000 + j0.0000  \\
\end{array} \right]
\\
\left[ T_V \right]_{old} = 
\left[ \begin{array}{cc}
0.0000 + j0.0000 & -0.7071 + j0.0000 \\
0.7057 - j0.7086 & -0.7071 + j0.0000 \\
\end{array} \right]
\\
\left[ Z_{cM} \right] = 
\left[ \begin{array}{cc}
85.3077 + j207.151 & 0.0000 - j0.0000 \\
0.0000 - j0.0000 & 36.4543 + j15.0707 \\
\end{array} \right] \Omega
\end{gathered}
\end{equation}
from which a number of observations may be made. Firstly, the modal propagation constants are in excellent agreement (within 0.03\% and 0.4\%) of HFSS, which gives
\begin{equation} \label{eq:Example3_HFSSGm}
\left[ \gamma_M \right] = 
\left[ \begin{array}{cc}
9.5405 + j23.0280 & 0.0000 + j0.0000 \\
0.0000 + j0.0000 & 16.5490 + j39.8950 \\
\end{array} \right] /m
\end{equation}
Secondly, the transformation matrices do not satisfy the current equalities, since the sum of the either of the positive or negative currents is not unity, and moreover, one entry of $\left[ T_V \right]$ has a significant imaginary component. Thirdly, the modal characteristic impedances differ greatly (44.8\% and 17.2\%, repsectively) from those given by HFSS:
\begin{equation} \label{eq:Example3_HFSSZcm}
\left[ Z_{cM} \right] = 
\left[ \begin{array}{cc}
292.5100 + j121.3000 & 0.0000 + j0.0000 \\
0.0000 + j0.0000 & 25.7750 + j10.6570 \\
\end{array} \right] /m
\end{equation}
The correction process yields the matrices
\begin{equation} \label{eq:Example3_CorrectedgIZcmTiTv}
\begin{gathered}
\left[ g_I \right] = 
\left[ \begin{array}{cccc}
1.4142 + j0.0000 & 0.0000 + j0.0000 \\
0.0000 + j0.0000 & 1.0000 + j0.0000 \\
\end{array} \right]
\\
\left[ T_V \right]_{new} = 
\left[ \begin{array}{cccc}
0.0000 + j0.0000 & 1.0000 + j0.0000 \\
1.0000 + j0.0000 & 1.0000 + j0.0000 \\
\end{array} \right]
\\
\left[ T_I \right]_{new} = 
\left[ \begin{array}{cccc}
-1.0000 + j0.0000 & 1.0000 + j0.0000 \\
1.0000 + j0.0000 & 0.0000 + j0.0000 \\
\end{array} \right]
\\\left[ Z_{cM} \right] = 
\left[ \begin{array}{cc}
292.7070 + j121.2430 & 0.0000 - j0.0000 \\
0.0000 - j0.0000 & 25.7771 + j10.6566 \\
\end{array} \right] \Omega
\end{gathered}
\end{equation}
from which it can be seen that the transformation matrices are real, and the characteristic impedance values are very close to those given by HFSS (0.03\% and smaller than 0.01\%), indicating that the modal transformation process proposed in this work is valid even in the presence of extreme loss.

\section{Conclusion}
\label{sec: Conclusion}

It has been shown that a unique relationship between voltage and current transformation matrices for TL modes can be extracted from the physical equivalence of power and current between domains, which represents an important improvement over previously proposed processes. Various examples further demonstrated the use and accuracy of the proposed correction process, through comparison with HFSS simulations. The unique determination of the properties of TL modes may find meaningful applications in fields such as electromagnetic compatibility, signal and power integrity in printed-circuit-board environments, and the analysis of coupling between closely spaced power system circuits.

\section*{Acknowledgment}
The authors would like to thank Mr. David Sawyer for inspirational discussion and proof-reading of this work.

\appendices

\section{Proof of Contra-Directed Phases}
\label{sec:AppendixContraDirectedPhases}

It is typically assumed that the forms of the transformation matrices $\left[T_V\right]$ and $\left[T_I\right]$ are not limited. However, this assumption implies that for a given excited mode, currents may be excited which possess any arbitrary phase, with respect to the excitation. Here, this possibility is formally investigated, assuming only the equivalence of power between the two domains.

Comparing (\ref{eq:G_Definition_1}) to the transpose of (\ref{eq:G_Definition_2}) in the case of corrected transformation matrices, and noting that the matrices $\left[Z_T\right]$ and $\left[Y_T\right]$ are symmetric, it can be concluded that
\begin{equation} \label{eq:Mode_Terminal_Propagation_Equated}
    \left[T_V\right]_{new} = \left[T_I\right]_{new}^{\text{-}1T} \left[D\right]
\end{equation}
where $\left[D\right]$ is an arbitrary matrix, which is required to be diagonal. Inserting $\left[T_V\right]_{new}$ into the proposed equality of power (\ref{eq:Modal_Terminal_Power_Condition}) yields 
\begin{equation} \label{eq:Diagonality_of_D}
    \left[D\right] = \overline{\left[T_I\right]}_{new}^{\text{\ -}1}  \left[T_I\right]_{new}
\end{equation}
The right-hand side of the preceding equation is not generally diagonal for any complex $\left[T_I\right]_{new}$, whereas $\left[D\right]$ is required to be diagonal. This apparent contradiction is resolved by rearranging (\ref{eq:Diagonality_of_D}) to give
\begin{equation} \label{eq:Diagonality_of_D2}
    \left[T_I\right]_{new} = \overline{\left[T_I\right]}_{new} \left[D\right]
\end{equation}
Observe that this can also be expressed as
\begin{equation} \label{eq:Diagonality_of_D3}
    \left[T_I\right]_{new} = \overline{\overline{\left[T_I\right]}_{new} \left[D\right]} \left[D\right] 
\end{equation}
Which can be further simplified to
\begin{equation} \label{eq:Diagonality_of_D4}
    \left[T_I\right]_{new} = \left[T_I\right]_{new} \left[\left|D\right|\right]^2
\end{equation}
This interesting result yields a pair of useful conclusions. Firstly, observing each column $n$ of the transformation matrix, where $\vec{T_I}_n = \vec{T_I}_n \left|D_n\right|^2$, it is noted that if $\left|D_n\right| \ne 1$, then $\vec{T_I}_n$ must equal $\vec{0}$, a case that can be physically discounted, since it represents a mode that excites no terminal-domain currents. Secondly, it is then noted that $D_n$ must be of the form $e^{j\theta_n}$, where $\theta_n$ is any real number. Inserting this observation into each row of (\ref{eq:Diagonality_of_D2}) yields 
\begin{equation} \label{eq:Angle_of_D}
    \vec{T_I}_n = \overline{\vec{T_I}}_n e^{j\theta_n}
\end{equation}
Generally, expressing each entry in $\vec{T_I}_n$ in polar form gives the following:
\begin{equation} \label{eq:Angle_of_D2}
    \left|{T_I}_{mn}\right|e^{j\phi_{mn}} = \left|{T_I}_{mn}\right|e^{-j\phi_{mn}} e^{j\theta_n} 
\end{equation}
where $\phi_{mn}$ is the phase angle of the complex entry ${T_I}_{mn}$. Note that this equation implies 
\begin{equation} \label{eq:Angle_of_D3}
    e^{j\phi_{mn}} = e^{j\left(\theta_n - \phi_{mn}\right)}
\end{equation}
and thus,
\begin{equation} \label{eq:Angle_of_D4}
    \phi_{mn} = \frac{\theta_n}{2} + i\pi,
\end{equation}
where $i$ is any integer. This result relates the phase of any element in a given column of $\left[T_I\right]_{new}$ to the value $\theta_n$, which only depends on the column $n$. That is to say, the elements of each column $n$ can only possess one of two possible phases, which are 180$^\circ$ out of phase with each other. A similar conclusion can be demonstrated for the phases of $\left[T_V\right]_{new}$.

\section{Proof of Real Transformation Matrices}
\label{sec:AppendixRealMatrices}

Having established that in the terminal domain, there exist only two phases of currents which are 180$^\circ$ apart, let the set $\left[T_I(+)\right]_{new}$ be defined as consisting of the terms of $\left[T_I\right]_{new}$ which possess a phase of $e^{j\theta_n}$, and $\left[T_I(-)\right]_{new}$ as consisting of the terms of $\left[T_I\right]_{new}$ which possess a phase of $e^{j\left(\theta_n + \pi\right)}$ (removing the $\pi$ phase shift):
\begin{subequations}
    \begin{equation} \label{eq:Conservation_of_Charge_Positive_Phases}
        \left[T_I(+)\right]_{new} = \left[|T_I(+)|\right]_{new} \left[e^{j\theta}\right]
    \end{equation}
    \begin{equation} \label{eq:Conservation_of_Charge_Negative_Phases}
        \left[T_I(-)\right]_{new} = \left[|T_I(-)|\right]_{new} \left[e^{j\theta}\right]
    \end{equation}
\end{subequations}
where $\left[|T_I(+)|\right]_{new}$ and $\left[|T_I(-)|\right]_{new}$ are matrices for which each entry is the magnitude of the corresponding entry in  $\left[T_I(+)\right]_{new}$ and $\left[T_I(-)\right]_{new}$, respectively, and $\left[e^{j\theta}\right]$ is a diagonal matrix, the entries of which are $e^{j\theta_n}$. Inserting these relations into (\ref{eq:Conservation_of_Charge_Positive3}) and (\ref{eq:Conservation_of_Charge_Negative3}), respectively, gives:
\begin{subequations}
    \begin{equation} \label{eq:Conservation_of_Charge_Positive_Phases2}
        1 = \vec{1}^T \left[|T_I(+)|\right]_{new} \left[e^{j\theta}\right] \vec{\delta}_n
    \end{equation}
    \begin{equation} \label{eq:Conservation_of_Charge_Negative_Phases2}
        1 = \vec{1}^T \left[|T_I(-)|\right]_{new} \left[e^{j\theta}\right] \vec{\delta}_n
    \end{equation}
\end{subequations}
which establish that $\theta_n = 2i\pi$. This is a remarkable result since, in conjunction with (\ref{eq:Modal_Terminal_Power_Condition}), it clearly indicates that the corrected transformation matrices are real.

\section{Proposed Correction Algorithm}
\label{sec:AppendixAlgorithm}

Obtaining the sum of each of the (non-reference) components of the columns of $\left[T_I\right]$ will yield two quantities: if these values are equal, then it can be stated conclusively that the current component on the reference conductor is zero, and that the value of $g_I$ is simply the inverse of either the positive or negative sum. If the two sums do not have the same magnitude, then the larger sum is the correct value (with the difference being the current component on the reference conductor), the inverse of which is $g_I$. With this knowledge, the pseudocode Alg. \ref{CorrectionAlgorithm} is proposed, where all scalar or matrix values are assumed to be of a complex, floating-point type, unless they are  indices.

This algorithm has an outermost loop which iterates over each column. For each column, a guard loop ensures that the column is sufficiently corrected before proceeding to the next column -- if it is not, then the inner loop is repeated. The inner loop has four main processes: in order of operation, the positive and negative sums are computed, and the larger sum is selected and evaluated. If the sum is sufficiently close to 1, the guard loop is notified that the process is complete, otherwise the column's $g_I$ is multiplied by the inverse of the sum and the loop is repeated. Lastly, once the complete $\left[g_I\right]$ matrix has been calculated for all columns, the corrected $\left[T_I\right]$ can be calculated, and utilizing the power equality, $\left[T_V\right]$ is calculated directly from this. Both corrected matrices are then returned.

\begin{algorithm}[t]
    \caption{$\left[T_V\right]$ and $\left[T_I\right]$ Correction Procedure}
    \label{CorrectionAlgorithm}
    \begin{algorithmic}[1]
        \Procedure{CorrectTvTi}{$\left[T_V\right],\left[T_I\right]$}
            \State $\left[g_I\right] \gets \left[I\right]$ \Comment{Initialize $\left[g_I\right]$ with identity}
            \For{\texttt{Each Column} n}
                \State $Balanced \gets false$ \Comment{True if sum is close to 1}
                \While{$Balanced = false$}
                    \State $NegSum \gets 0$
                    \State $PosSum \gets 0$
                    \State $LargerSum \gets 0$
                    \For{\texttt{Each Row} k}
                    \State $a \gets T_I\left(k,n\right) \times g_I\left(n,n\right)$
                        \If{$\Re(a) \leq$ 0} 
                            \State $NegSum \gets NegSum - a$
                        \Else
                            \State $PosSum \gets PosSum + a$
                        \EndIf
                    \EndFor
                    \If{$\Re(NegSum) \geq \Re(PosSum)$} 
                        \State $LargerSum \gets NegSum $
                    \Else
                        \State $LargerSum \gets PosSum $
                    \EndIf
                    \If{$\left|\Re\left(LargerSum\right) - 1\right| \leq$ 0.001} 
                        \State $Balanced \gets true$
                    \Else
                        \State $g_I\left(n,n\right) \gets g_I\left(n,n\right) \div LargerSum$
                    \EndIf
                \EndWhile
            \EndFor
            \State $\left[T_I\right] \gets \left[T_I\right] \times  \left[g_I\right]$
            \State $\left[T_V\right] \gets {\left[T_I\right]^{-1}}^{T}$
            \State \textbf{return} $\left(\left[T_V\right],\left[T_I\right]\right)$
        \EndProcedure
    \end{algorithmic}
\end{algorithm}


\bibliographystyle{IEEEtran}

\begin{IEEEbiography}[{\includegraphics[width=0.98in]{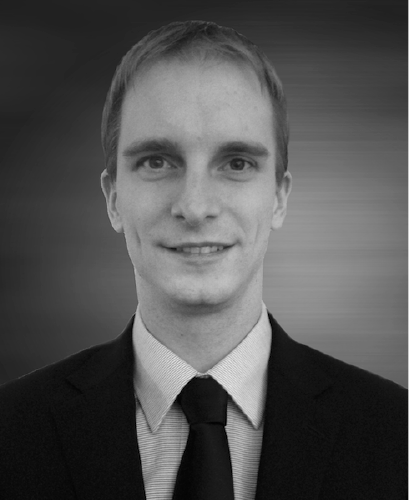}}]{Stuart Barth}
\noindent (S'07–-GSM'11) received the B.Sc. and M.Sc. degrees in electrical engineering in 2012 and 2015, respectively, from the University of Alberta, Edmonton, AB, Canada where he is currently working towards the Ph.D. degree.
          		
His current research interests include the study of multiconductor transmission-line RF/microwave circuits, dispersion engineering of periodic structures, fundamental electromagnetic theory, and antenna radiation-pattern shaping.
          		
Mr. Barth received the IEEE AP-S Pre-Doctoral Research Award in 2014, and the IEEE AP-S Doctoral Research Grant in 2016 for his ongoing research into electromagnetic bandgap structures for antenna and waveguide applications. He serves as an Officer of the IEEE Northern Canada Section MTT-S/AP-S joint chapter.
\end{IEEEbiography}
          	
          	
\begin{IEEEbiography}[{\includegraphics[width=0.98in]{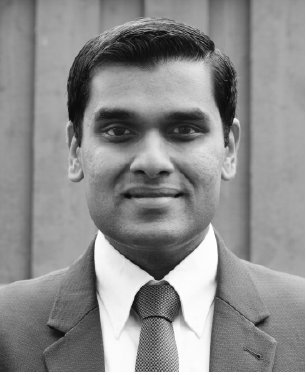}}]{Ashwin K. Iyer}
\noindent (S'01--M'09--SM'14) received the B.A.Sc. (Hons.), M.A.Sc., and Ph.D. degrees in electrical engineering from the University of Toronto, Toronto, ON, Canada, in 2001, 2003, and 2009, respectively, where he was involved in the dis- covery and development of the negative-refractive- index transmission-line approach to metamaterial design and the realization of metamaterial lenses for free-space microwave subdiffraction imaging.
He is currently an Associate Professor with the Department of Electrical and Computer Engineering, University of Alberta, Edmonton, AB, Canada, where he leads a team of graduate students investigating novel RF/microwave circuits and techniques, fundamental electromagnetic theory, antennas, and engineered metamaterials, with an emphasis on their applications to microwave and optical devices, defense technologies, and biomedicine. He has co-authored a number of highlycited papers and four book chapters on the subject of metamaterials.
Dr. Iyer is a Registered Member of the Association of Professional Engineers and Geoscientists of Alberta (APEGA). He was a recipient of several awards, including the 2008 R. W. P. King Award and the 2015 Donald G. Dudley Jr. Undergraduate Teaching Award, presented by the IEEE AP-S. His students are also the recipients of several major national and international awards for their research. He serves as a Co-Chair of the IEEE Northern Canada Section’s joint chapter of the AP-S and MTT-S societies. Since 2012, he has served as an Associate Editor of the IEEE TRANSACTIONS
ON ANTENNAS AND PROPAGATION.
\end{IEEEbiography}





\end{document}